\documentstyle[12pt,a4wide]{article}
\begin{document}
\title{\bf Electrostatics in a Schwarzschild black hole pierced by
a cosmic string}
\author{B. Linet \thanks{E-mail: linet@celfi.phys.univ-tours.fr} \\ 
\small Laboratoire de Math\'ematiques et Physique Th\'eorique \\
\small CNRS/UPRES-A 6083, Universit\'e Fran\c{c}ois Rabelais \\
\small Parc de Grandmont 37200 TOURS, France}
\date{}
\maketitle
\thispagestyle{empty}

\begin{abstract}
We explicitly determine the expression of the electrostatic potential 
generated by a point charge at rest in the Schwarzschild black hole
pierced by a cosmic string. We can then calculate the electrostatic 
self-energy. From this, we find again the upper entropy bound for
a charged object by employing thermodynamics of the black hole.
\end{abstract}

\section{Introduction}

It is well known for a long time that a point charge at rest in a static spacetime 
feels an electrostatic self-force. The calculation is performed by considering
the global electrostatic potential determined as the solution of the
Maxwell equations in the background metric of the spacetime. However, it would
seem that its existence was a curiosity. The situation has recently undergone a
change when Bekenstein and Mayo \cite{bek1}
and Hod \cite{hod} have derived the upper entropy bound for a charged object by 
requiring the validity of thermodynamics of the Reissner-Nordstr\"{o}m 
black hole. Their proof takes really into account the expression of the 
electrostatic self-energy for a point charge at rest in a Schwarzschild 
black hole which has been previously determined in closed form 
\cite{smi1,zel,lea1}.

The purpose of this work is to extend these results to a new case where it is possible
to determine explicitly the electrostatic self-energy. We consider the
spacetime, introduced by Aryal {\em et al} \cite{ary}, which
describes a Schwarzschild black hole pierced by a cosmic string. 
It represents a straight cosmic string, infinitely thin, passing through a 
spherically symmetric black hole. It is
obtained by cutting a wedge in the Schwarzschild geometry. So, 
in the coordinate system $(t,r,\theta ,\varphi )$ with $0\leq \varphi <2\pi$,
the metric can be written
\begin{equation}\label{Sch}
ds^2=-\left( 1-\frac{2m}{r}\right) dt^2+\left( 1-\frac{2m}{r}\right)^{-1}dr^2
+r^2d\theta^2+B^2r^2\sin^2\theta d\varphi^2 
\end{equation}
where $m$ is a positive parameter and $B$ is related to the linear mass density 
$\mu$ of the cosmic string by $B=1-4\mu$ with $0<B<1$. We only consider
the spacetime outside the horizon, i.e. for $r>2m$.

In section 2, we summarise the Maxwell equations in metric (\ref{Sch})
and, in the case $1/2<B<1$, we give explicitly the expression of the electrostatic
potential generated by a point charge at rest. The
proof that this expression obeys the electrostatic equation is fulfilled in section 3. 
Taking into account the found expression of the electrostatic self-energy,  
we derive in section 4 the entropy bound for a charged object by employing 
thermodynamics of the black hole. We add some concluding remarks in
section 5.

\section{Electrostatic potential}

The Maxwell equations in metric (\ref{Sch}), having as source a point 
charge $e$ located at the position $(r_0,\theta_0,\varphi_0)$ with $r_0>2m$, 
reduce to
\begin{equation}\label{Max}
\partial_i\left( \sqrt{-g}F^{i0}\right) =-\frac{e}{4\pi}
\delta (r-r_0)\delta (\theta -
\theta_0)\delta (\varphi -\varphi_0) \quad {\rm with}\quad 
F_{i0}=\partial_iA_0
\end{equation}
where $A_0$ is the electrostatic potential. 
According to (\ref{Max}), the electrostatic equation for $A_0$ can be written
\begin{eqnarray}\label{elect} 
\nonumber & & \frac{1}{r^2}\frac{\partial}{\partial r}\left( r^2\frac{\partial}{\partial r}
A_0\right) +\frac{1}{r(r-2m)\sin \theta}\frac{\partial}{\partial \theta}
\left( \sin \theta \frac{\partial}{\partial \theta}A_0\right)
+\frac{1}{B^2r(r-2m)}\frac{\partial^2}{\partial \varphi^2}A_0 \\
& & =-\frac{e}{4\pi Br^2\sin \theta}\delta (r-r_0)\delta (\theta -\theta_0)
\delta (\varphi -\varphi_0) \, .
\end{eqnarray}
We point out that the application of the Gauss theorem to 
equation (\ref{Max}) yields
\begin{equation}\label{gauss}
A_0(r,\theta ,\varphi )\sim \frac{e}{Br} \quad {\rm as}\quad 
r\rightarrow \infty  
\end{equation}
if there is no electric flux through the horizon.
Furthermore, we require that the electromagnetic field is regular at the 
horizon by imposing that $F^{\mu \nu}F_{\mu \nu}$ is finite as $r\rightarrow 2m$.

We limit ourselves to the case where $1/2<B<1$ which is physically justified 
for a cosmic string since $\mu \ll 1$. Without loss of generality, 
we put $\varphi_0=\pi$ to simplify. The expression of the electrostatic potential $A_0$
satisfying equation (\ref{elect}) with the desired boundary conditions can
be expressed as the following sum
\begin{equation}\label{ssc}
A_0(r,\theta ,\varphi )=V^*(r,\theta ,\varphi )+
V_B(r,\theta ,\varphi )+\frac{em}{Brr_0}
\end{equation}
where the expressions of $V^*$ and $V_B$ are given below. 

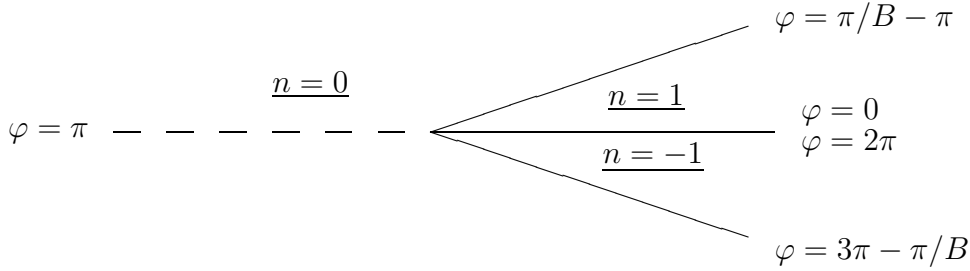
\begin{figure}
\begin{picture}(320,100)(10,10)
\put(10,58){$\varphi =\pi$}
\put(50,60){\line(1,0){10}}\put(70,60){\line(1,0){10}}\put(90,60){\line(1,0){10}}
\put(110,60){\line(1,0){10}}\put(130,60){\line(1,0){10}}\put(150,60){\line(1,0){10}}
\put(170,60){\line(3,1){120}}
\put(300,100){$\varphi =\pi /B-\pi$}
\put(300,12){$\varphi =3\pi -\pi /B$}
\put(310,65){$\varphi =0$}\put(310,53){$\varphi =2\pi$}
\put(170,60){\line(3,-1){120}}
\put(170,60){\line(10,0){130}}
\put(237,70){\underline{$n=1$}}\put(235,48){\underline{$n=-1$}}
\put(110,75){\underline{$n=0$}}
\end{picture}
\caption{Regions delimited by the hypersurfaces $\varphi =$ constant}
\end{figure}

To express the potential $V^*$, we must consider the regions of the spacetime
delimited by the hypersurfaces $\varphi =$ constant as shown on Figure~1. 
We have
\begin{equation}\label{vetoile}
V^*(r,\theta ,\varphi )=\left\{ \begin{array}{ll}
V_C(r,\sigma_0(\theta ,\varphi )]+V_C[r,\sigma_1(\theta ,\varphi )] 
& \quad 0<\varphi <\pi /B -\pi \\
V_C[r,\sigma_0(\theta ,\varphi )] & \quad \pi /B-\pi <\varphi <3\pi -\pi /B \\
V_C[r,\sigma_0(\theta ,\varphi )]+V_C[r,\sigma_{-1}(\theta ,\varphi )] 
& \quad 3\pi -\pi /B<\varphi <2\pi 
\end{array} \right.
\end{equation}
where $V_C$ is the Copson potential \cite{cop} which is a solution to 
electrostatic equation (\ref{elect}) with $B=1$, i.e. for the
Schwarzschild black hole. Its expression is
\begin{equation}\label{copson}
V_C[r,\sigma ]=\frac{e}{rr_0}\frac{(r-m)(r_0-m)-m^2\sigma}
{[(r-m)^2+(r_0-m)^2-m^2-2(r-m)(r_0-m)\sigma +m^2\sigma^2]^{1/2}}
\end{equation}
The variables $\sigma_n$ in formula (\ref{vetoile})
are the following functions of $\theta$ and $\varphi$ 
\begin{eqnarray}\label{sig0}
\nonumber & & \sigma_0(\theta ,\varphi )=\cos \theta \cos \theta_0+\sin \theta
\sin \theta_0\cos B(\varphi -\pi ) \, , \\
& & \sigma_1(\theta ,\varphi )= \cos \theta \cos \theta_0+\sin \theta
\sin \theta_0\cos B(\varphi +\pi ) \, , \\
\nonumber & & \sigma_{-1}(\theta ,\varphi )=\cos \theta \cos \theta_0+\sin \theta 
\sin \theta_0 \cos B(\varphi -3\pi ) \, .
\end{eqnarray}

The potential $V_B$ is given by the integral expression
\begin{equation}\label{vb}
V_B(r,\theta ,\varphi )=\frac{1}{2\pi B}\int_{0}^{\infty}
V_C[r,k(\theta ,x)]F_B(\varphi ,x)dx
\end{equation}
where the function $k$ is given by
\begin{equation}\label{kx}
k(\theta ,x)=\cos \theta \cos \theta_0-\sin \theta \sin \theta_0\cosh x
\end{equation}
and the function $F_B$ by
\begin{equation}\label{fb}
F_B(\varphi ,x)=-\frac{\sin (\varphi -\pi /B)}
{\cosh x/B+\cos (\varphi -\pi /B)}
+\frac{\sin (\varphi +\pi /B)}{\cosh x/B+\cos (\varphi +\pi /B)} \, .
\end{equation}

In the Schwarzschild black hole, sum (\ref{ssc}) with
$V_1=0$ and $V^*=V_C$ yields the electrostatic potential that we have
already obtained \cite{lin1}.  On the other hand for the 
cosmic string, i.e. $m=0$, we find our previous result \cite{lin2}, already
known in the case of a wedge in flat space \cite{gar,obe}.

The electrostatic self-potential in a neighbourhood of the point charge is 
$A_0(r,\theta ,\varphi )-V_C[r,\sigma_0(\theta ,\varphi )]$. In 
consequence, the electrostatic self-energy $W_{self}$ is
\begin{equation}\label{self}
W_{self}(r_0,\theta_0)=\frac{e^2m}{2Br_{0}^{2}}
-\frac{e\sin \pi/B}{2\pi B}\int_{0}^{\infty}
V_C[r_0,k(\theta_0,x)]\frac{dx}{\cosh x/B-\cos \pi /B} \, .
\end{equation}
From (\ref{self}), we can deduce the electrostatique self-force which has
been already obtained in the Schwarzschild black hole \cite{smi1,zel,lea1} 
and in the cosmic string \cite{lin3,smi2}.

\section{Checking of the electrostatic solution}

We must firstly verify that sum (\ref{ssc}) is a solution to 
equation (\ref{elect}). The potential $V^*$ is obviously a local solution 
since the Copson potential $V_C$ expressed in variables $r$, $\theta$ 
and $\phi$ with $\phi =B\varphi$ obeys the electrostatic equation for the 
Schwarzschild black hole. As a consequence, the function
$V_C[r,k(\theta ,x)]$ satisfies
\[
\frac{1}{r^2}\frac{\partial}{\partial r}\left( r^2\frac{\partial}{\partial r}
V_C\right)+\frac{1}{r(r-2m)\sin \theta}\frac{\partial}{\partial \theta}
\left( \sin \theta \frac{\partial}{\partial \theta}V_C\right)
-\frac{1}{r(r-2m)}\frac{\partial^2}{\partial x^2}V_C=0 \, .
\]
Then, according to its expression (\ref{vb}), potential $V_B$ obeys 
electrostatic equation (\ref{elect}) without second member if we have
\[
\int_{0}^{\infty}\frac{\partial^2}{\partial x^2}V_C[r,k(\theta ,x)]
F_B(\varphi ,x)dx+\frac{1}{B^2}\int_{0}^{\infty}V_C[r,k(\theta ,x)]
\frac{\partial^2}{\partial \varphi^2}F_B(\varphi ,x)dx =0 \, .
\]
Now from expression (\ref{fb}), we verify that
\begin{equation}\label{resc}
\frac{\partial^2}{\partial x^2}F_B(\varphi ,x)+\frac{1}{B^2}
\frac{\partial^2}{\partial \varphi^2}F_B(\varphi ,x)=0
\end{equation}
which ensures the condition mentioned above after two
sucessive integrations by part.
We notice that the potential $em/Brr_0$ is a homogeneous solution 
to electrostatic equation which is regular at the horizon.

We secondly check that sum (\ref{ssc}) is continuous. At $\varphi =0$,
it is clear because $V^*(r,\theta ,0)=V^*(r,\theta ,2\pi )$ since
$\sigma_1(r,\theta ,0)=\sigma_{-1}(r,\theta ,2\pi )$.
At $\varphi =\pi /B-\pi$, we introduce $\epsilon$ by setting
$\varphi =\pi /B-\pi +\epsilon$ and then the potential $V_B$ becomes
\begin{eqnarray}\label{epsilon}
\nonumber & & V_B(r,\theta ,\pi /B-\pi +\epsilon )=
\frac{\sin \epsilon}{2\pi B}\int_{0}^{\infty}
V_C[r,k(\theta,x)]\frac{dx}{\cosh x/B -\cos \epsilon} \\
& & -\frac{\sin (2\pi /B+\epsilon )}{2\pi B}\int_{0}^{\infty}V_C[r,k(\theta ,x)]
\frac{dx}{\cosh x/B-\cos (2\pi /B+\epsilon)} \, .
\end{eqnarray}
We write down the following integral
\[
\sin \epsilon \int_{0}^{x}\frac{dy}{\cosh y-\cos \epsilon}=
2\arctan \left( \tanh \frac{x}{2}\cot \frac{\epsilon}{2}\right) 
\quad {\rm with}\quad \epsilon \neq 0 \, .
\]
By integrating by part the first term of expression (\ref{epsilon}), we get
\[
\frac{1}{2}V_C[r,k(\theta ,\infty )]-\frac{1}{\pi}\int_{0}^{\infty}
\frac{\partial}{\partial y}V_C[r,k(\theta ,By)]\arctan \left( 
\tanh \frac{y}{2}\cot \frac{\epsilon}{2}\right) dy \, .
\]
But the function $\arctan$ is bounded by $\pi /2$ and consequently we may
take the limit $\epsilon \rightarrow 0$ inside the integral. We obtain thereby
\[
\frac{1}{2}V_C[r,k(\theta ,\infty )]-\frac{1}{2}\left\{ 
V_C[r,k(\theta ,\infty )]-V_C[r,k(\theta ,0)]\right\} 
\quad {\rm as}\quad \epsilon \rightarrow 0  \; \epsilon >0 \, .
\]
We have thus prove that integral expression (\ref{epsilon}) verifies 
\begin{eqnarray}\label{limite}
\nonumber & & \lim_{\epsilon \rightarrow 0\; \epsilon >0}V_B
(r,\theta ,\pi /B-\pi +\epsilon )=\frac{1}{2}V_C[r,k(\theta ,0)] \\
& & -\frac{\sin 2\pi /B}{2\pi B}\int_{0}^{\infty}
V_C[r,k(\theta ,By)]\frac{dy}{\cosh y-\cos 2\pi /B}
\end{eqnarray}
and another with $\epsilon <0$ yielding $-V_C/2$ in formula (\ref{limite}).
On the other hand, the potential $V^*$ verifies
\begin{equation}\label{disc}
\lim_{\epsilon \rightarrow 0 \; \epsilon >0} \left(
V^*(r,\theta ,\pi /B-\pi +\epsilon )-V^*(r,\theta ,\pi /B-\pi -\epsilon )
\right) = -V_C[r,k(\theta ,0)] \, .
\end{equation}
By combining results (\ref{limite}) and (\ref{disc}), we thus obtain  that
the potentials $V_B$ and $V^*$ are both discontinuous at 
$\varphi =\pi /B -\pi$ whereas their sum is regular. Of course, 
the potential $V^*+V_B$ is also continuous at $\varphi =3\pi -\pi /B$ by symmetry. 
In conclusion, the electrostatic potential $A_0$ is a smooth function only
singular at the position of the point charge.
Furthermore, it is easy to show
that the derivative of the electrostatic potential $A_0$ with
respect to $\varphi$ is everywhere continuous.

We thirdly determine the asymptotic form of the electrostatic potential $A_0$. 
From expression (\ref{vetoile}) of $V^*$, we have immediately 
\begin{equation}\label{v*asymp}
V^*(r,\theta ,\varphi )\sim \left\{ \begin{array}{ll}
2e(r_0-m)/rr_0 & \quad 0<\varphi <\pi /B -\pi \\
e(r_0-m)/rr_0 & \quad \pi /B -\pi <\varphi <3\pi -\pi /B \quad 
{\rm as}\quad r\rightarrow \infty \\
2e(r_0-m)/rr_0 & \quad 3\pi -\pi /B <\varphi <2\pi \end{array} \right. \, .
\end{equation}
On the other hand, from expression (\ref{vb}) of $V_B$ we get 
\begin{equation}\label{vbasymp}
V_B(r,\theta ,\varphi )\sim
\frac{e(r_0-m)}{rr_0}g(\varphi ) \quad {\rm as}\quad r\rightarrow \infty
\end{equation}  
with
\[
g(\varphi )=\frac{1}{2\pi}\int_{0}^{\infty}\left[
\frac{\sin (\varphi +\pi /B)}{\cosh x+\cos (\varphi +\pi /B)}
-\frac{\sin (\varphi -\pi /B)}{\cosh x+\cos (\varphi -\pi /B)}\right] dx
\]
which can be integrated by elementary methods
\begin{equation}\label{gdyexp}
g(\varphi )= \left\{ \begin{array}{ll}
1/B-2 & \quad 0<\varphi < \pi /B-\pi \\
1/B-1 & \quad \pi /B-\pi <\varphi < 3\pi -\pi /B \\
1/B-2 & \quad 3\pi -\pi /B <\varphi <2\pi \end{array} \right. \, .
\end{equation}
By using (\ref{v*asymp}) and (\ref{vbasymp}) with (\ref{gdyexp}), we obtain that
the electrostatic potential (\ref{ssc}) has the desired asymptotic form (\ref{gauss}).

At last, we must verify that the electromagnetic field derived from the 
electrostatic potential (\ref{ssc}) is regular at the horizon. 
This point results of the fact that $V_C$ tends to $e/rr_0$ when
$r\rightarrow 2m$.

\section{Entropy bound for a charged object}

We now consider the spacetime which describes a Reissner-Nordstr\"{o}m
black hole pierced by a cosmic string. It is obtained by cutting a wedge
in the Reissner-Nordstr\"{o}m geometry. In the coordinate
$(t,r,\theta ,\varphi )$ with $0\leq \varphi <2\pi$, the metric can be written
\begin{eqnarray}\label{RN}
\nonumber & & ds^2=-\left( 1-\frac{2E}{Br}+\frac{q^2}{B^2r^2}\right) dt^2 \\
& & +\left( 1-\frac{2E}{Br}+\frac{q^2}{B^2r^2}\right)^{-1}dr^2+
r^2d\theta^2+B^2r^2\sin^2\theta d\varphi^2 
\end{eqnarray}
where $E$ and $q$ are two parameters. We only consider the spacetime
outside the outer horizon, i.e. $r>(E+\sqrt{E^2-q^2})/B$ by assuming that
$E^2>q^2$. Following \cite{ary,mar}, we interpret $E$ as the energy of 
the black hole. Clearly, $q$ is the electric charge of the black hole.
For $q=0$, metric (\ref{RN}) reduces to metric (\ref{Sch})
by setting $m=E/B$. The horizon area ${\cal A}$ of the black hole defined by 
metric (\ref{RN}) has the expression
\begin{equation}
{\cal A}(E,q)=\frac{4\pi}{B}\left( E+\sqrt{E^2-q^2}\right)^2 
\end{equation}
and the entropy $S_{BH}$ of the black hole is given by
\begin{equation}\label{aire}
S_{BH}(E,q)=\frac{1}{4}{\cal A}(E,q) \, .
\end{equation}

The Reissner-Nordstr\"{o}m black hole pierced by a cosmic string linearised
with respect to its electric charge $q$ is described by metric (\ref{Sch})
plus an electromagnetic test field having the electrostatic potential
\begin{equation}\label{ext}
A_{0}^{ext}(r,\theta ,\varphi )=\frac{q}{Br} \, .
\end{equation}
Moreover, the black hole entropy (\ref{aire}) reduces to
\begin{equation}\label{entropy}
S_{BH}(E,q) \approx \frac{2\pi}{B} \left( 2E^2-q^2\right) \, .
\end{equation}

The original method of Bekenstein \cite{bek2} for finding the entropy bound
for a neutral object in the Schwarzschild black hole has been recently 
extented for charged object in the
Reissner-Nordstr\"{o}m black hole \cite{bek1,hod}. Referring to
\cite{bek1,hod}, we
recall that the energy ${\cal E}$ of a charged object with a mass $\mu$, an 
electric charge $e$ and a radius $R$ located at the position 
$(r_0,\theta_0)$ in metric (\ref{Sch}), in presence of the exterior 
electrostatic potential (\ref{ext}), has the expression
\begin{equation}\label{energy}
{\cal E}=\sqrt{1-\frac{2E}{Br_0}}+\frac{eq}{Br_0}+
W_{self}(r_0,\theta_0)
\end{equation}
where $W_{self}$ is the electrostatic self-energy (\ref{self}). 
When the charged object is just outside the horizon, 
its energy (\ref{energy}), for a very small proper length $R$, is
\begin{equation}\label{last}
{\cal E}_{last}\sim \frac{\mu RB}{4E}+\frac{eq}{2E}+W_{self}(2E/B,\theta_0)
\quad {\rm as}\quad R\rightarrow 0  \, .
\end{equation}
In this state, the system formed by the black hole and the charged object has
an entropy $S_{BH}(E,q)+S$ where $S$ is
the entropy of the charged object. When the charged object falls in the 
horizon, the final state is a Reissner-Nordstr\"{o}m black hole with the new parameters
\begin{equation}
E_f=E+{\cal E}_{last} \quad {\rm and}\quad q_f=q+e \, .
\end{equation}
But in this final state, the entropy is $S_{BH}(E_f,q_f)$. We now write down
the generalised second law of thermodynamics 
\begin{equation}\label{ineg}
S_{BH}(E_f,q_f)\geq S_{BH}(E,q)+S \, .
\end{equation}

We can calculate $\triangle S_{BH}=S_{BH}(E_f,q_f)-S_{BH}(E,q)$ from 
expression (\ref{entropy}). We keep only linear terms in ${\cal E}_{last}$.
By this way, we thus exclude a possible
gravitational self-force which should be quadratic in $\mu$
as in a cosmic string \cite{smi2}. We find
\begin{equation}\label{triangle}
\triangle S_{BH}=\frac{4\pi}{B} \left[2E{\cal E}_{last}-eq-
\frac{e^2}{2} \right]  \, .
\end{equation}
By inserting (\ref{last}) into (\ref{triangle}), we get
\begin{equation}\label{bound}
\triangle S_{BH}=\frac{4\pi}{B}\left[ \frac{\mu RB}{2}+2E\left( 
\frac{e^2}{8E}+\frac{e^2B}{8E}g(\pi ) \right) 
-\frac{e^2}{2} \right] \, .
\end{equation}
where $g(\pi )=1/B-1$ by formula (\ref{gdyexp}).
According to inequality (\ref{ineg}), we obtain then from (\ref{bound})
the desired entropy bound
\begin{equation}
S \leq 2\pi \left[ \mu R-\frac{e^2}{2}\right] \, ,
\end{equation}
initially derived by Zaslavskii \cite{zas} in another context.

\section{Conclusion}

We have determined the explicit expressions of the electrostatic potential
and self-energy in the Schwarzschild black hole pierced by a cosmic string.
We can extend our method to the static, spherically symmetric spacetimes 
pierced by a cosmic string when the electrostatic potential is known 
in absence of a cosmic string: Brans-Dicke \cite{lin4}
and Reissner-Nordstr\"{o}m \cite{lea2}.

We have found again the upper entropy bound for a charged object
by employing thermodynamics of the Reissner-Nordstr\"{o}m black hole 
pierced by a cosmic string. To prove this, we have used the value of the 
electrostatic self-energy at the horizon of the Schwarzschild black hole 
pierced by a cosmic string. This result confirms the physical importance of 
the electrostatic self-force.

\end{document}